\documentclass[manuscript,nonacm]{acmart}

\usepackage{tikz}
\usetikzlibrary{arrows.meta, positioning, shapes.misc}

\usepackage{framed}
\usepackage{xcolor}
\definecolor{framecolor}{rgb}{0.8,0.2,0.2} % 定义框体颜色（红色）
 {\endMakeFramed}

\AtBeginDocument{%
  }

\setcopyright{acmlicensed}
\copyrightyear{2026}
\acmYear{2026}
\acmDOI{XXXXXXX.XXXXXXX}

\acmConference[Conference acronym 'XX]{}{}{}
\acmISBN{978-1-4503-XXXX-X/18/06}

\begin{document}

%%
%% The \textit{"title"} command has an optional parameter,
%% allowing the author to define a \textit{"short title"} to be used in page headers.
\title[Research IDE]{Probing the Future of Meta-Analysis: Eliciting Design Principles via an Agentic Research IDE}
%%
%% The \textit{"author"} command and its associated commands are used to define
%% the authors and their affiliations.
%% Of note is the shared affiliation of the first two authors, and the
%% \textit{"authornote"} and \textit{"authornotemark"} commands
%% used to denote shared contribution to the research.

\author{Sizhe Cheng}
\email{sizhe003@e.ntu.edu.sg}
\affiliation{%
  \institution{College of Computing and Data Science}
  \institution{Nanyang Technological University}
  \city{Singapore}
  \country{Singapore}
}

\author{Feng Liang}
\email{feng011@e.ntu.edu.sg}
\affiliation{%
  \institution{College of Computing and Data Science}
  \institution{Nanyang Technological University}
  \city{Singapore}
  \country{Singapore}
}

\author{Yuhan Wen}
\email{G250004@e.ntu.edu.sg}
\affiliation{%
  \institution{College of Computing and Data Science}
  \institution{Nanyang Technological University}
  \city{Singapore}
  \country{Singapore}
}

\author{Xipei Yu}
\email{yuxipei120@Gmail.com}
\affiliation{%
  \institution{Independent Researcher}
  \city{Guangzhou}
  \country{China}
}

\author{Yong Wang}
\email{yong-wang@ntu.edu.sg}
\affiliation{%
  \institution{College of Computing and Data Science}
  \institution{Nanyang Technological University}
  \city{Singapore}
  \country{Singapore}
}

% \author{Ben Trovato}
% \authornote{Both authors contributed equally to this research.}
% \email{trovato@corporation.com}
% \orcid{1234-5678-9012}
% \author{G.K.M. Tobin}
% \authornotemark[1]
% \email{webmaster@marysville-ohio.com}
% \affiliation{%
%   \institution{Institute for Clarity in Documentation}
%   \city{Dublin}
%   \state{Ohio}
%   \country{USA}
% }

%%
%% By default, the full list of authors will be used in the page
%% headers. Often, this list is too long, and will overlap
%% other information printed in the page headers. This command allows
%% the author to define a more concise list
%% of authors' names for this purpose.
\renewcommand{\shortauthors}{B. Sun et al.}

%%
%% The abstract is a short summary of the work to be presented in the
%% article.

\begin{abstract}
Meta-analyses and systematic reviews demand rigorous abductive reasoning to build, test, and refine hypotheses across vast, heterogeneous literature. While NLP advancements have automated parts of this pipeline, existing tools often detach researchers from the cognitive loop or function merely as retrieval engines, leading to loss of intellectual ownership and frequent context switching. We present Research IDE, a prototype reimagining authoring environments through the "Research as Code" metaphor. Research IDE embeds a multi-agent backend into the writing flow, enabling in-situ verification via "hypothesis breakpoints." A one-week field deployment with 8 domain experts, followed by a reflective workshop, as a Research through Design (RtD) probe, reveals that users strongly preferred this verification workflow, actively leveraged prior knowledge for confirmation, and reported that breakpoints sparked insights. Drawing from participant feedback and suggestions, we derive design implications for future AI-assisted research tools that fully preserve researcher autonomy and intellectual ownership while harnessing computational scale. 
\end{abstract}

%%
%% The code below is generated by the tool at http://dl.acm.org/ccs.cfm.
%% Please copy and paste the code instead of the example below.
%%

\begin{CCSXML}
<ccs2012>
   <concept>
       <concept_id>10003120.10003121</concept_id>
       <concept_desc>Human-centered computing~Human computer interaction (HCI)</concept_desc>
       <concept_significance>500</concept_significance>
   </concept>
   <concept>
       <concept_id>10003120.10003121.10003124</concept_id>
       <concept_desc>Human-centered computing~Interactive systems and tools</concept_desc>
       <concept_significance>500</concept_significance>
   </concept>
\end{CCSXML}

\ccsdesc[500]{Human-centered computing~Human computer interaction (HCI)}
\ccsdesc[500]{Human-centered computing~Interactive systems and tools}

%%
%% Keywords. The author(s) should pick words that accurately describe
%% the work being presented. Separate the keywords with commas.
\keywords{AI-assisted research, systematic review, large language model, human-AI collaboration}

%% A \textit{"teaser"} image appears between the author and affiliation
%% information and the body of the document, and typically spans the
%% page.
% \begin{teaserfigure}
%   \includegraphics[width=\textwidth]{fig/figure_01.png}
%   \caption{Test}
%   \label{fig:teaser}
% \end{teaserfigure}

% \received{20 February 2007}
% \received[revised]{12 March 2009}
% \received[accepted]{5 June 2009}

%%
%% This command processes the author and affiliation and title
%% information and builds the first part of the formatted document.
\maketitle

\section{Introduction}

Systematic reviews and meta-analyses demand complex abductive reasoning, which involves forming explanatory hypotheses from evidence and iteratively revising them as new findings emerge, requiring researchers to iteratively propose hypotheses, seek evidence, and refine interpretations across vast, heterogeneous literatures~\cite{greenhalgh2005storylines, snyder2016identifying}. This cognitively intensive sensemaking process~\cite{page2021prisma, borah2017analysis} has attracted growing interest in AI assistance; however, effective support remains elusive due to the holistic nature of the workflow.

Current LLM-powered tools exhibit a fundamental interaction paradigm mismatch. While recent frameworks have evolved from evaluating GPT-4's drafting capabilities~\cite{wang2024using} to automating synthesis via RAG~\cite{han2024automating}, these approaches often enforce rigid linearity. Specialized tools (e.g., Elicit, Scim~\cite{fok2023scim}) target isolated sub-tasks like retrieval or reading, neglecting the unified workflow. Even advanced multi-agent systems~\cite{qiu2025completing} decouple verification from writing, treating humans as passive supervisors. This fragmentation erodes intellectual ownership, as researchers are forced to accept automated outputs rather than actively engaging in the logic construction essential to original scholarship~\cite{Blaizot2022Using}.

To address this, we propose ``Research as Code'', a metaphor that reframes hypothesis verification as analogous to code debugging. We instantiate this vision in \textbf{Research IDE}, a prototype featuring Hypothesis Breakpoints: interactive markers enabling researchers to pause and verify claims against literature in-situ, supported by a multi-agent backend for cross-document reasoning.

Following Research through Design (RtD)~\cite{zimmerman2007research}, we deployed Research IDE with 8 domain experts for one week, followed by a reflective workshop. Findings reveal users shifted from passive search toward active falsification, deliberately challenging their hypotheses. Importantly, participants reported that the breakpoint mechanism preserved their sense of intellectual ownership while sparking insights. In summary, we contribute: (1) the ``Research as Code'' metaphor and Hypothesis Breakpoint mechanism and Research IDE as a functional probe; (2) empirical insights on expert verification behaviors; and (3) design implications for AI-assisted research tools balancing factual accuracy with researcher autonomy.

\section{Research IDE: The Technology Probe}

We developed Research IDE, a web-based writing environment that instantiates the "Research as Code" metaphor to bridge the gap between authoring and verification.

\begin{figure*}[h!]
\centering
\includegraphics[width=\linewidth]{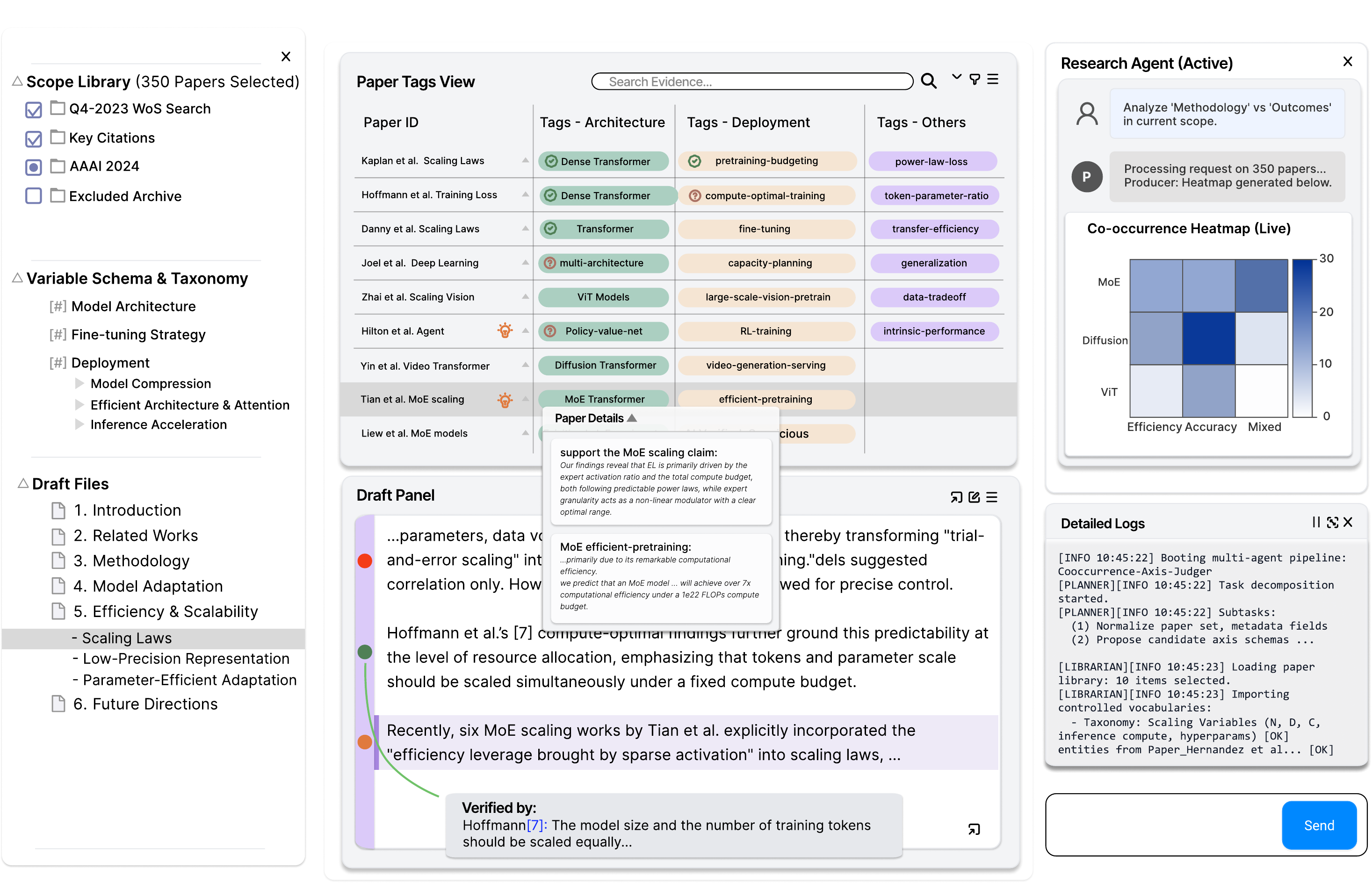}
\caption{Interaction Design of Research IDE. The interface embeds verification directly into the authoring flow. As illustrated, setting a Hypothesis Breakpoint (e.g., the red indicator) seamlessly triggers the Multi-Agent Backend. The orchestration layer then dispatches specialized agents to retrieve and synthesize evidence from the literature, returning grounded insights to the user without requiring context switching. Users can ask agents questions at any time for clear visual explanations.}
\label{fig:figure2}
\end{figure*}

\subsection{Design Goals}
Research IDE was designed to explore how AI can support rigorous abductive reasoning without diminishing researcher agency. Based on challenges identified in prior work, we defined the following design goals for our probe.

\textbf{DG1: Preserve Intellectual Ownership.} Instead of replacing the author with generative text, the system should support users in their roles as the primary thinkers and reasoners of ideas, ensuring they maintain cognitive control over the narrative.

\textbf{DG2: Enable In-Situ Verification.} To minimize the cognitive load caused by context switching, the tool must integrate evidence retrieval and verification directly within the writing environment.

\textbf{DG3: Support Active Falsification over Retrieval.} Moving beyond simple keyword search, the system should treat user claims as hypotheses to be tested, aligning with the reasoning and verification process.

\subsection{System Implementation}
We implemented Research IDE as a functional, high-fidelity prototype to ensure ecological validity during field deployment. The system achieves our design goals through a frontend React web page driven by an asynchronous multi-agent backend.

\textbf{Writing as Debugging Interface.} To enable in-situ verification (DG2), we introduce the ``Hypothesis Breakpoint'' mechanism. As researchers draft, they can insert a breakpoint, effectively pausing to test their logic. The sidebar then renders evidence and categorizes them into Supporting (Green), Skeptical (Orange) or Contradicting (Red) that prioritizes active falsification (DG3). Crucially, the system prioritizes high controllability (DG1): users can configure verification parameters by defining custom taxonomy prompts or scoping the search to specific paper sets. Furthermore, the interface exposes the backend's reasoning processes in real-time to foster trust.

\textbf{Agentic Collaboration.} The verification pipeline is orchestrated by a multi-agent framework that synthesizes architectures from prior literature~\cite{wang2024using,han2024automating,qiu2025completing}. It consists of four specialized agents powered by OpenAI's gpt5-mini: (1) A Planner decomposes the user's high-level claim into queryable sub-questions; (2) A Librarian executes RAG to search evidence over the user-defined scope; (3) A Reasoner performs cross-document analysis to identify consensus or dissensus; and (4) A Producer presents these insights to users and visualize them to drive evidence on the front end.

% \section{Study Design}
% % 1.25 pages（N=12 需要写清楚，否则 reviewer 会质疑）
% \subsection{Participants}
% % N=12，背景（是否有混合会议经验、性别/年龄范围），是否成对/成组
% \subsection{Design and Conditions}
% % 关键：你是“单条件体验研究”还是“对照研究”？
% % Poster建议：单条件 + 角色对比（remote vs co-located）仍然保留
% \subsection{Procedure}
% % 简要：pre -> icebreaker -> meeting task -> post -> interview
% \subsection{Measures}
% % 建议分两类：
% % (1) Quant: 社会存在感/连接感/开场尴尬/愿意发言/隐私担忧/打扰感（Likert）
% % (2) Qual: 半结构访谈 + 主题分析
% \subsection{Analysis}
% % Quant: 描述统计 +（可选）配对检验（remote vs co-located）/效应量
% % Qual: thematic analysis（简述编码流程）

\section{User Study: Field Deployment and Reflection}

We adopt a Research through Design approach~\cite{zimmerman2007research} and position the current system as a design probe to provoke user reflection. Our goal is to deploy the system into real-world workflows to validate the probe while inspiring users to reflect on the tension between AI automation and human agency, thereby deriving requirements for future system design.

\subsection{Participants and Environment}
We recruited 8 domain experts ($P1-P8$, 3 females, 5 males; ages 23-31) from diverse research backgrounds, including Human-Computer Interaction, Visualization, Artificial Intelligence, and Quantum Computing. All participants are active researchers and currently have an ongoing need for literature reviews. The study was conducted in a real-world setting to ensure ecological validity. Participants used Research IDE on their personal laptops and were asked to use it as a verification tool for an ongoing writing task.

\subsection{Procedure}
The study consisted of two phases spanning one week:

\textbf{Phase 1: Longitudinal Deployment (1 Week).} 
After a 20-minute tutorial, participants used the system asynchronously for 7 days. We encouraged them to use the \textit{Hypothesis Breakpoint} feature whenever they felt uncertain about a claim in their draft. This distinct period allowed participants to move beyond the novelty effect and experience the friction and benefits of active falsification in real practice. During the 7 days, in the backend logs, a total of 105 user breakpoints were recorded and verified, and 548 papers were assigned labels.

\textbf{Phase 2: Collaborative Reflective Workshop (90 mins).} 
After the deployment, we conducted a workshop. A device with Research IDE open and diverse screening recordings prepared by the authors are available for users to use and demonstrate at any time to facilitate discussions in conjunction with specific functions. The workshop was organized by two authors, and participants were divided into 2 groups of 4 people each. We structured the session into three stages:

\begin{enumerate}
    \item \textbf{Situation Reconstruction via Theme Cards (20 mins).} 
    We presented participants with Theme Cards and Emoji Cards representing specific cognitive states (e.g., \textit{``A moment of distrust,''} \textit{``A surprise finding,''} \textit{``A conflict with the Agent''}). 
    Participants were asked to select one card and pair it with a specific artifact from the mocked typical usage logs (e.g., a screenshot of a specific verification card or a chat history) to reconstruct the context. They then shared the background story and the significance of that specific interaction.

    \item \textbf{Trust Calibration Discussion (50 mins).} 
    Building on the shared incidents, we facilitated a discussion on the alignment between their trust in GenAI and the probe's design. We examined whether the Research IDE's current features (e.g., source visualization, traffic-light indicators) matched their trust requirements at different stages of the workflow, from retrieval to synthesis. We specifically probed for trust gaps, moments where the system asked for trust but the user hesitated. In each trust gap, researchers freely discussed and combined the feature cards and agent cards pre-prepared by the authors to envision new features.

    \item \textbf{Opportunity Voting (20 mins).} 
    In the final stage, we synthesized the discussed friction points into actionable design requirements. Participants categorized requirements into Must-haves vs. Nice-to-haves, helping us distill the final design principles.
\end{enumerate}

\subsection{Data Analysis}
We recorded and transcribed the workshop session. We applied reflexive Thematic Analysis~\cite{braun2006using} to the transcripts. Two authors independently coded the data, moving from semantic codes to latent themes. The findings presented below focus on the emergent design implications for agentic writing support.

% \section{Results}
% % 1.5 pages（核心升级点：N=12 可以“更像结果”但仍要紧凑）
% \subsection{Quantitative Results}
% % 用1张小表或1张图：均值+置信区间/误差条
% % 报告 remote vs co-located 差异（若你有within-subject）
% \subsection{Qualitative Themes}
% % 2-3个主题，每个主题1-2句解释 + 1句代表性引用（总引用控制在2-4条）
% % 推荐主题：
% % T1: co-creation降低开场压力/给远端“合法发言入口”
% % T2: 心跳具身反馈提升对远端的注意与在场感
% % T3: 解释与隐私张力（可见性、被误读、职业场景顾虑）

\section{Findings}

Synthesizing data from two distinct workshop groups (Group A \& B, Total $N=8$), our analysis revealed four overarching themes regarding how domain experts envision AI and Research IDE in their workflow.

\subsection{Theme 1: AI as Iterator, Not Author}
A strong consensus emerged regarding the role of AI. Participants firmly rejected the previous AI Author or AI Co-author mode, citing a loss of agency and divergent thinking. Instead, they praised our AI agents as high-speed iterators.
\begin{itemize}
    \item \textbf{Importance of fine-granularity Verification:} Users from both groups praised and emphasized the verification behavior of Research IDE, particularly how breakpoints can directly link to the corresponding original content. Both groups of users created designs to further enhance this to build trust, with observed patterns including multi-model voting and paper citation graphs integrated with clearer implementations. 
    \item \textbf{The 80\% Threshold:} Surprisingly, users were tolerant of imperfection in exploratory tasks. For complex tasks like Taxonomy Building, most participants agreed that an accuracy of 70-80\% was acceptable. They value AI's ability to validate multiple schemes for ``low-cost trial and error'', knowing they can discard it without cost if it fails. Users generally agreed that this rapid iteration can continuously ``stimulate their own thinking'' and dynamically refine the methodology for paper classification, as ``a most difficult writing part in a review.''
    \item \textbf{The ``Drag-to-Cite'' Controversy:} In Group A's discussion regarding the automatic generation of review sentences by dragging and dropping papers, a moment of sharp disagreement emerged. While P1 prioritized efficiency, P3 strongly objected: \textit{``If AI writes the connecting sentences, I lose the logical flow. I want to insert citation tags by dragging papers, but I must write the verbs myself.''} This clearly draws a boundary: some users accept \textbf{narrative automation}, while some users can only accept \textbf{functional automation} (linking, formatting).
    \item \textbf {Different Workflows.} Interestingly, while fine granularity is a common requirement, users have many disagreements based on their accustomed workflows. Disagreements include whether the evidence chain should be represented in the form of text, images, or knowledge graphs. Another point of contention is whether evidence should be placed next to the writing area to support a real-time verification workflow, or in a separate interface to support a "verify all at once" mode, based on their vastly different review habits.
\end{itemize}

\subsection{Theme 2: The Integration Holy Grail}
Participants generally believe that Research IDE has the potential to bridge current siloed workflows. A recurring need is to achieve the seamless integration of the three current stages: writing, management, and verification.
\begin{itemize}
    \item \textbf{The Cost of Context Switching:} During the co-design activity, participants consistently agreed the AI agent should be in a collapsible sidebar, ``just like the VSCode Copilot design''. In addition, they articulated a trade-off between screen space and accessibility, preferring to keep the AI accessible via a drag-and-drop interaction. 
    \item \textbf{User Notes as Ground Truth:} A critical insight was the hierarchy of context. Participants in Group A argued that their own annotations in PDF readers should be treated as the ``Ground Truth'' for the AI. P4 noted: \textit{``If I highlighted it in Zotero, the AI should know it's important. I shouldn't have to prompt it again.''}
    \item \textbf{Implicit Ground Truth Construction:} Group B echoed on this, meanwhile proposing views on tags. They proposed that interaction history is vital for sensemaking. P6 described a vision of ``Flowing Tags'': \textit{``As I converse, it should keep adding tags by my casual instructions... eventually clustering literature into a natural atmosphere.''} 
\end{itemize}

\subsection{Theme 3: Structure Begets Trust}
When discussing trust calibration, participants rejected unstructured text generation. They indicated that \textbf{Structure} is the primary vehicle for explainability.
\begin{itemize}
    \item \textbf{Pre-structuring Input:} Participants suggested that trust begins at the input level. They believed that if the AI explicitly identifies the structural components of a paper (e.g., parsing ``Methods'' vs. ``Conclusion'') before processing, its output is inherently more reliable. in Group A, P2 remarked: \textit{``I trust it more if I know it's reading the 'Methodology' section when I ask about sample size.''}
    \item \textbf{The Paradox of Freedom:} This need for structure was reinforced by Group B's criticism of overly flexible components, such as direct dialogue with the agent. P7 noted a problem with current chat-based tools like: \textit{``It's too free... The AI doesn't understand my intent unless I prompt it perfectly.''} This confirms that in exploratory research, structural scaffolding is preferred over infinite freedom.
    \item \textbf{Natural Structure.} Group B rejected the binary choice between structured folders (too rigid) and flat search (too messy). Instead, they proposed a dynamic clustering model. P7 described a vision where manual folders provide the skeleton, while flowing taxonomy derived from chat history provide the muscle. This creates a natural structure where papers are organized by both their origin (Folder) and their semantic context (Tag).
\end{itemize}

% \section{Discussion and Implications}
% % 0.75 page
% \subsection{Design Implications for Hybrid Icebreakers}
% % 2-3条可迁移的设计启示（短句）
% \subsection{Design Implications for Biosignal Sharing}
% % “可选/分级/抽象化/情境化”这种总结
% \subsection{Limitations and Next Steps}
% % 2-3句话：样本、实验环境、规模化、多设备识别等

\section{Discussion and Implications}

Our findings challenge the prevailing chatbot or oracle paradigm in academic AI tools. By summarizing the discussions in the workshop, we explore three design implications.

\textbf{Redefining Agency.}
Users provide a critical nuance to the concept of human-AI collaboration. They do not always reject automation but they reject loss of agency. Just as a code linter flags syntax errors without rewriting the algorithm, the future tools should flag logical gaps, cite missing evidence, or suggest stylistic polish, while strictly refraining from usurping the authorial voice.
These tools should strictly separate functional automation (formatting, linking, polishing) from epistemic automation (logic generation) to meet different agency needs.

\textbf{Scaffolding over Freedom.}
The paradox of freedom observed in the workshop reveals that the open-ended interaction model fails in exploratory research. Unlike coding, where intent is often explicit, scientific inquiry involves ambiguous knowledge gaps. 
Future tools should move away from an oracle. Instead, they must provide structural scaffolding—visualizing the topology of evidence and offering constrained entry points to reduce decision fatigue.

\textbf{From Static Retrieval to Emergent Structure.}
Participants' rejection of siloed workflows and manual tagging points to a new model of knowledge management. Their design choices suggest that metadata should not be a prerequisite chore, but a byproduct of sensemaking.
We envision a system where ground-truth tags is constructed implicitly. As users converse about specific topics or highlight text in Zotero, the system should treat these interactions as evolving metadata, automatically structuring and clustering the literature library in the background.

\section{Limitations and Future Work}
Our study employs a technology probe with a small sample of domain experts ($N=8$). While this yielded deep qualitative insights, the relatively small sample sizes may affect data reliability. Also, the sample is limited to researchers in the STEM field, which weakens the generalizability of the study. Additionally, we have observed that users have vastly different research workflows and different agency definition and needs, but we have not yet explored in depth how to design tools to support all these workflows. Future work should investigate more user, including more iterations and testing, to satisfy more requirements.

\section{Conclusion}

Systematic reviewing and meta-analyses are cornerstones of scientific progress, yet it remains a bottleneck in the research workflow. In this work, we introduced Research IDE, a technology probe that reimagines the authoring environment through the lens of "Research as Code."
Our workshops revealed that domain experts do not seek an AI that writes for them, but one that argues with them. They value traceability over creativity, and structure over freedom.
The "Research as Code" metaphor successfully bridged the gap between authoring and verification, suggesting that for high-stakes cognitive tasks, users prefer a tool that acts as a rigorous debugger of thoughts rather than a generative co-author. We hope this work inspires a new generation of human-centered tools that prioritize productive friction and epistemic agency, ensuring that AI serves to sharpen, rather than replace, human scientific reasoning.

\begin{acks}
    Test
\end{acks}

% \clearpage

\bibliographystyle{ACM-Reference-Format}
\bibliography{reference}

\end{document}